\documentstyle[12pt]{article}

\def\beq{\begin{equation}}
\def\eeq{\end{equation}}
\def\beqa{\begin{eqnarray}}
\def\eeqa{\end{eqnarray}}
\def\dprod{\displaystyle\prod}

\def\dfrac{\displaystyle\frac} 
\def\aa{\alpha}

\def\V{\cal V}

\def\b0{\bar 0}
\def\a'{\alpha^{'}}
\def\V'{V^{'}}
\def\bV'{\bar \V'}
\begin{document}
\topmargin 0pt
\oddsidemargin 1mm
\begin{titlepage}
\begin{flushright}
NORDITA-98/77 HE\\
DSF-TH-98/46\\
December 1998\\
\end{flushright}
\setcounter{page}{0}
\vspace{15mm}
\begin{center}
{\Large Prescriptions for Off-Shell Bosonic String Amplitudes~\footnote{Talk 
delivered by one of the authors (R. M.)
at the Mid-Term TMR meeting, Corf\`u (Greece), 20-26 September 1998.
Work partially supported by the European
Commission TMR programme ERBFMRX-CT96-0045 (Nordita-Copenhagen and
Universit\`a di Napoli).}}
 
\vspace{20mm}

{\large Luigi Cappiello}, {\large Antonella Liccardo},
{\large Roberto Pettorino}, {\large Franco Pezzella~\footnote{e-mail:
Name.Surname@na.infn.it}}\\
{\em Dipartimento di Scienze Fisiche, Universit\`{a} di Napoli\\
and I.N.F.N., Sezione di Napoli,\\
Mostra d'Oltremare, pad. 19, I-80125 Napoli, Italy}\\
{\large Raffaele Marotta~\footnote{e-mail: marotta@nbivms.nbi.dk}}\\
{\em Nordita, Blegdamsvej 17, DK-2100 Copenhagen \O, Denmark}
\end{center}
\vspace{7mm}

\begin{abstract}
We give, in the framework of the bosonic string theory, simple prescriptions 
for computing, at tree and one-loop levels, off-shell string amplitudes 
for open and closed string massless states.
In particular we obtain a tree amplitude for three open strings
that in the field theory limit coincides with the three-gluon vertex in
the usual covariant gauge and two-string one-loop amplitudes 
satisfying the property of transversality.
\end{abstract}

\vfill

\end{titlepage}

\newpage
Perturbative calculations in string theories involve scattering amplitudes 
with on-shell physical external states.
The underlying conformal (and superconformal) invariance 
imposes that vertex operators have to be primary fields 
and this condition constrains the external states to be on-shell. 

However, on-shell conditions, even if very important for the 
internal consistency of the theory, sometimes turn out to be a serious 
limitation in practical calculations. 

One of these limitations appears in the field theory limit or zero slope limit 
($\alpha^{'} \rightarrow 0$) of string amplitudes.

It is well-known that in the zero-slope limit the various string models have 
to reproduce 
perturbative aspects of ordinary gauge field theories of the
fundamental interactions, including gravity. 

In particular, closed strings can be used to get information 
on perturbative quantum gravity and its divergences \cite{BDS}, 
while open strings, with a $SU(\mbox{N})$ 
gauge symmetry embodied by Chan-Paton factors, give informations 
on non-abelian gauge theories. Indeed, instead of computing field theory 
amplitudes by conventional methods, which are known to be algebraically 
very complex expecially at high loop order and in quantum gravity, one can 
compute the corresponding string amplitudes, that have more compact 
expressions and are much fewer, and then evaluate their low-energy limit
\cite{MT}$\div$\cite{BK1}.

In field theory, however, one is also interested in calculating off-shell, 
gauge dependent quantities such as anomalous dimensions or general Green 
functions. It is then clear that a reliable prescription for off-shell
string amplitudes has to be formulated.

Furthermore, once an off-shell prescription has been defined, one can 
also use it to compute processes involving 
interactions among D-branes \cite{P,RT,BS}. Therefore, there are
many reasons of interest in studying off-shell string amplitudes.

Off-shell extensions have been studied in great deal 
until now 
\cite{CF}$\div$ \cite{DMLRM} and it can be shown that while on-shell  
string amplitudes are independent from the choice of local coordinate systems 
defined around the punctures \cite{DPFHLS,NW}, their off-shell 
extension depends on how that choice is performed.
In the following, in the framework of the operatorial formalism, we will give
a general prescription for computing off-shell string amplitudes at 
tree and one-loop levels. We will examine the general $M$-string amplitude
for $M$ massless states 
and we will propose a simple choice of local coordinates at the 
punctures on the 
Riemann surface, such that $none$ of the on-shell conditions on the external
states has to be kept. 
The tree three-gluon 
amplitude obtained with this procedure, reproduces, in the low energy limit, 
the usual three-gluon vertex written in the usual covariant gauge while the one-loop 
string amplitudes obtained in the case $M=2$ turn out to satisfy the 
transversality condition.

Finally, we would like also to stress that in the following it is not necessary 
to use a consistent model 
of string expecially if the final aim is the exploration of the field theory 
limit. In fact, as shown in Ref. \cite{B}, in order to extract field 
theory results from 
string theories one can use the bosonic model and disregard by hand tachyons
in calculations. Keeping this in mind we will consider the simplest, but
rich enough for further extensions, the bosonic string model and, by using the 
operatorial formalism, we will obtain off-shell bosonic string amplitudes.

The starting point of the operatorial formalism we use, both in closed and 
open string, 
is the $M$-string $g$-loop Vertex \cite{DPFHLS} $V_{M;g}$, which can be 
considered as a 
generating functional for scattering amplitudes among arbitrary string 
states, at all order of the perturbative expansion.
In fact, by saturating the operator $V_{M;g}$ on $M$ external states 
$|\aa_1>,\cdots, |\aa_2>$, one obtains the corresponding amplitude:

\beq
A^g(\aa_1,\cdots\aa_M)=V_{M;g}|\aa_1>,\cdots,|\aa_2> .
\label{gam}
\eeq

Let us restrict to the  case of external massless 
states of the bosonic closed string. Such states depend both on the 
polarization tensor $\epsilon$, that we decompose   
as $\epsilon_{\mu \nu} = \xi_{\mu} \otimes \bar{\xi}_{\nu} $, and on the 
momentum $p$: 

\begin{equation}
|\epsilon; p>={\cal N}_0\epsilon _{\mu \nu}\aa _{-1}^\mu \bar{\aa }%
_{-1}^\nu |p> . \label{MS0}
\end{equation}

where ${\cal N}_0=\kappa /\pi $, with $\kappa$ being the gravitational 
 constant in $d$ dimensions \cite{CMPP1}.
 
The state defined in (\ref{MS0}), according to the symmetry properties of the polarization tensor
$\epsilon_{\mu \nu}$, describes the antisymmetric tensor or a combination 
of gravitons and dilatons.

We first consider the one-loop case.
When the operator $V_{1,M}$ is saturated on $M$ massless states (\ref{MS0}),
it gives the following amplitude \cite{CMPP1}:
\[
A_{M;1}\, =\frac{1}{(2\pi \a')^{d/2}}\! \int [ {\mbox d} m]^{1}_{M}
\sqrt{\frac{\a'}{2}} p^{(i)} \cdot
\left.\! \left[ \sqrt{\frac{\a'}{2}} p^{(i)} + {\alpha}_{1}^{(i)} 
\partial_{z}
+ \bar{\alpha}_{1}^{(i)} \partial_{\bar{z}} \right] \log |V'_{i}(z)|^2
\left.\right|_{z=0} \right\}
\]
\[
\times \exp \bigg\{ \sum_{\stackrel{i,j=1}{i \neq j}}^{M} 
\left[ \sqrt{\frac{\a'}{2}} p^{(i)} + \xi^{(i)} V'_{i}(0)
\partial_{z_{i}} + \bar{\xi}^{(i)} \bar{V}'_{i}(0) \partial_{\bar{z}_{i}}
\right]
\]
\[
\cdot  \left[ \sqrt{\frac{\a'}{2}} p^{(j)} + \xi^{(j)}
V'_{j}(0\partial_{z_{j}} + \bar{\xi}^{(j)} \bar{V}'_{j}(0)
\partial_{\bar{z}_{j}} \right]  {\cal G}(z_{i},z_{j} ) \bigg\}
\]
\beq
\times \exp \left\{  -2 \sum_{i=1}^{M} \xi^{(i)} \cdot 
\bar{\xi}^{(i)} |V'_{i}(0)|^{2} \partial_{z} \partial_{\bar{z}} 
\,\, {\cal G}(z,z_{i} )\left.  \right|_{{z=z_{i}}} \right\} .
\label{AML1}
\end{equation}

This expression has to be understood as an expansion in the
polarization vectors $\xi^{(i)}$
and $\bar{\xi}^{(i)}$ restricted only to the linear terms in each 
of them.

The one-loop measure is \cite{DFLS}

$$
\left[ {\mbox d}m \right]_M^{1}=\prod_{i=2}^{M}
\frac{d^2 z_i}{|V'_{i}(0)|^{2}}\,\dfrac{d^2k}{
|k|^4}\left[ -\ln |k|\right] ^{-d/2\,}\dprod\limits_{n=1}^{+\infty }\left(
\left| 1-k^n\right| ^2\right) ^{2-d} 
$$

where $k$ is the multiplier of the projective 
transformation and we have used the projective invariance to fix $z_1=1$
\cite{DPFHLS}.

${\cal G}(z_i,z_j)$ is the one-loop Green function given by \cite{GSW}:
 
$$
{\cal G}(z_i,z_j)=\frac{1}{2} \log  \left| (z_{i} - z_{j}) 
\dprod\limits_{n=1}^{+\infty }\frac{%
(z_i-k^nz_j)(z_j-k^nz_i)}{z_iz_j(1-k^n)^2}\right| ^2+
\frac{1}{2} \frac{ {\log}{}^2 |z_1/z_2| }{\log |k|} .
$$

Finally the functions $V_i(z)$ are local conformal coordinates defined 
around the punctures and satisfying the constraint 

\beq
V_i(0)=z_i
\label{G1}
\eeq

When considered on-shell, i.e. when  $p^2=0$ and 
$p\cdot \xi=p\cdot\bar{\xi}=0$, the amplitude (\ref{AML1})
does not depend on the $V_i$'s. If we relax these conditions we get 
off-shell string amplitudes depending on the choice of the local 
coordinates defined around the punctures. 
This is somewhat analogous to what happens 
in gauge theories, where on-shell amplitudes are gauge invariant, while 
their off-shell counterpart are not.

We can use the freedom of choosing the conformal local map $V_i(z)$ in order 
to write the amplitude (\ref{AML1}) in a more simple and compact form 
depending only on the Green function.

This can be obtained first rescaling the Green function as follows:  

\beq
G(z_{i}, z_{j}) =  {\cal G}(z_{i},z_{j}) - \frac{1}{4} \log |V'_{i}(0)
V'_{j}(0)|^{2}      
\eeq

This expression coincides with the one given in literature \cite{M,RS},
where the conformal maps $V_i$'s depend on the moduli of the world- sheet 
and on the punctures.

Furthermore, by choosing at one-loop:
  
\beq
V'_{i}(0)= z_{i} \label{G2}   
\eeq

we reproduce the translational invariant Green function on the 
torus \cite{GSW}.
However the condition (\ref{G2}) is not sufficient to write  
the amplitude (\ref{AML1}) only in terms of the Green function; this can be
achieved imposing the further condition:

\beq
V''_{i}(0)=z_{i}  \label{G3}
\eeq

The constraints (\ref{G1}), (\ref{G2}) and (\ref{G3}) may be satisfied,
choosing, for instance, the following holomorphic local coordinate map
at the puncture $z_{i}$ :
\beq
V_{i}(z) = z_{i} e^{z} .  \label{GAUGE}
\eeq
This choice allows us to rewrite the eq. (\ref{AML1}) as:
 
\[
A_{M;1}\, =\, \frac{1}{ (2 \pi\,\a')^{d/2} } \! \int [ {\mbox d} m]^{1}_{M}
\exp \bigg\{ \sum_{\stackrel{i,j=1}{i \neq j}}^{M} 
\left[ \sqrt{\frac{\a'}{2}} p^{(i)} + \xi^{(i)} V'_{i}(0)
\partial_{z_{i}} + \bar{\xi}^{(i)} \bar{V}'_{i}(0) \partial_{\bar{z}_{i}}
\right]
\]
\[ 
\cdot  \left[ \sqrt{\frac{\a'}{2}} p^{(j)} + \xi^{(j)}
V'_{j}(0)\partial_{z_{j}} + \bar{\xi}^{(j)} \bar{V}'_{j}(0)
\partial_{\bar{z}_{j}} \right]  G(z_{i},z_{j} ) \bigg\}
\]

\beq
\times \exp \left\{  -2 \sum_{i=1}^{M} \xi^{(i)} \cdot 
\bar{\xi}^{(i)} |V'_{i}(0)|^{2} \partial_{z} \partial_{\bar{z}} 
\,\, G(z,z_{i} )\left.  \right|_{{z=z_{i}}} \right\} .
\label{AML2}
\end{equation}
and it reproduces, for small values of $z$,
the gauge 
\beq
V_{i}(z) = z_{i}z + z_{i}      \label{GAUGEOP}
\eeq
that has been proposed for the open string~\cite{DMLRM}, 
together with the {\em semi off-shell} conditions $p^2 \neq 0$ and
$\epsilon \cdot p = 0$ for external photons. 

We point out that, instead, our choice does not need to be coupled to 
{\em any} 
other  condition. Our proposal for the local maps $V_{i}(z)$'s includes 
{\em all} the off-shell conditions.

In the following we will check our prescription for off-shell bosonic
string amplitudes by considering the case $M=2$.

The one-loop two-string amplitude is obtained by expanding $A_{2;1}$
up to terms linear in each external polarization. After some algebra we get:

\[
A_{2;1} = \frac{{\cal N}_{0}^{2}}{(2\pi \,\a')^{d/2}} \epsilon^{\mu \nu (1)}\epsilon^{\rho \sigma (2)}
                          T_{\mu \nu \rho \sigma}
\]
with

$$
\begin{array}{ll}
T_{\mu \nu \rho \sigma} =  
  &  4 \bigg[ \, \eta _{\mu \nu }\eta _{\rho \sigma }\ a_1+\eta _{\mu
\sigma }\eta _{\nu \rho }\ a_2+\eta _{\mu \rho }\eta _{\nu \sigma }\ a_3
\\  
& +\alpha ^{\prime } \bigg( \, \eta _{\mu \rho }\ p^{(1)}_{\sigma }\ p^{(2)}_{\nu \ }a_4 - \eta
_{\mu \nu }\ p^{(1)}_{\rho }\ p^{(1)}_{\sigma \ }a_5+\eta _{\mu \sigma }\
p^{(1)}_{\rho
}\ p^{(2)}_{\nu \ }a_6 \\  
& + \eta _{\nu \rho }\ p^{(1)}_{\sigma }\ p^{(2)}_{\mu \ }a_7-\eta _{\rho 
\sigma }\
p^{(2)}_{\mu }\ p^{(2)}_{\nu \ }a_8 + \eta _{\nu \sigma }\ p^{(1)}_{\rho }\ 
p^{(2)}_{\mu \
}a_9 \bigg) \\    
& + (\alpha ^{\prime })^2\ p^{(1)}_{\rho }\ p^{(1)}_{\nu \ }\
p^{(2)}_{\mu }\ p^{(2)}_{\sigma \ }a_{10} \bigg] 
\end{array}
$$

where the explicit expressions for the coefficients $a_i$'s, $i=1,\ldots,10$, 
depending on the Green
function and its derivative, can be found in the Ref. \cite{CMPP1}.

By integrating by parts all the terms containing double derivatives of the
Green function and using the symmetry in the exchange of the two external 
states, it can be shown that only three $a_{i}$'s are independent
\cite{CMPP1}.

Rewriting $T_{\mu \nu \rho \sigma}$ in terms of the 
independents coefficients, this expression drastically simplifies
reducing to:

\beqa
T_{\mu \nu \rho \sigma} &  =  & \bigg\{ - \frac{2}{p^{2}} (a_{3} + a_{2} )
\left[ \eta_{\mu \rho} p_{\nu} p_{\sigma} + \eta_{\nu \rho} p_{\mu} p_{\sigma}
+ \eta_{\mu \sigma} p_{\rho} p_{\nu} + \eta_{\nu \sigma} p_{\rho} p_{\mu}
\right]  \nonumber \\
& {} & - \frac{4}{p^{2}} a_{1} \left[ \eta_{\mu \nu} p_{\rho} p_{\sigma} +
\eta_{\rho \sigma} p_{\mu} p_{\nu} \right] + \frac{4}{p^{4}} (a_{1}+a_{2}
+a_{3}) p_{\mu} p_{\nu} p_{\rho} p_{\sigma} \nonumber \\
& {} & +  2 (a_{3}+a_{2}) \left[ \eta_{\mu \rho} \eta_{\nu \sigma} +
\eta_{\nu \rho} \eta_{\mu \sigma} \right] + 4 a_{1} \eta_{\mu \nu}
\eta_{\rho \sigma} \bigg\} \nonumber \\
& {} & + \bigg\{ - \frac{2}{p^{2}} (a_{3} - a_{2} ) \left[ \eta_{\mu \rho}
p_{\nu} p_{\sigma} - \eta_{\nu \rho} p_{\mu}p_{\sigma} + \eta_{\nu \sigma}
p_{\mu} p_{\rho} - \eta_{\mu \sigma} p_{\rho} p_{\nu} \right] \nonumber \\
& {} & + 2 (a_{3} - a_{2} )  \left[ \eta_{\mu \rho} \eta_{\nu \sigma}
- \eta_{\mu \sigma} \eta_{\nu \rho} \right] \bigg\} \ \equiv \ S_{\mu \nu
\rho \sigma} + A_{\mu \nu \rho \sigma} 
\eeqa

where we used the momentum conservation for setting $p^{(1)}=-p^{(2)} \equiv p$
and made explicit the symmetry properties on the indices $(\mu \nu)$ which
refer to the polarization tensor of the particle (1)  and 
$(\rho \sigma)$ which refer to the one of the particle (2).

It is simple to check that the one-loop amplitude, 
{\em as an off-shell string amplitude}, has the  
property of being transverse:

\[
p^{\mu} T_{\mu \nu \rho \sigma} = 0  .
\]

Let us to check that our prescription also works in the open string case
\cite{DMLRM}. In this case we will show that our procedure gives consistent 
off-shell amplitudes both at tree level and one-loop.

The starting point is again the  $M$-gluon $g$-loop amplitude \cite{CMPP1},
computed saturating the $M$-string $g$-loop Vertex for the open string
on $M$ states so defined:
\beq
| \epsilon; p > = N_0^{ph.} \epsilon_{\mu} \alpha^{\mu}_{-1} |p>
\label{OMS}
\eeq
where $N_0^{ph.}= g_d \sqrt{2 \a'}$ being $g_d$ the gauge coupling constant.

The corresponding amplitude is given by
\footnote{The relation between the string coupling constant $g_s$ and 
$g_d$ is given by \cite{DMLRM}: 
$g_s=\frac{g_d}{2}(2\a')^{1-d/4}$}:

\[
A_{M;g} =\frac{\mbox{N$^g$Tr}[\lambda^{a_1}\cdots\lambda^{a_M}]}{
(2\pi)^{dg}(2\a')^{d/2}}g_s^{2g-2}
\int [ {\mbox d} m]^{g}_{M}
\]
\[
\times \exp \left\{\frac{1}{2} \sum_{i=1}^{M} \sqrt{2 \a'} p^{(i)}
\left. \left[ \sqrt{2 \a'} p^{(i)} + \xi_i 
\partial_{z} \right] \log V'_{i}(z)
\right|_{z=0}\! \right\}
\]
\begin{equation}
\times \exp \bigg\{\!\frac{1}{2} \sum_{\stackrel{i,j=1}{i \neq j}}^{M}\! 
\left[ \sqrt{2 \a'} p^{(i)} + \xi_i  V'_i(0)
\partial_{z_{i}}\right]
\cdot \left[ \sqrt{2 \a'} p^{(j)} + \xi_j  V'_j(0)
\partial_{z_{j}} \right] {\cal G}(z_{i},z_{j} ) \bigg\}
\label{VMPO}
\end{equation}
where $N^g\,$Tr$[\lambda^{a_1}\cdots\lambda^{a_n}]$ is the Chan-Paton factor 
appropriate for an $M$-gluon g-loop planar diagram, with $\lambda$'s  
being the generators of $SU(\mbox{N})$ in the fundamental tachyons, and
\beq
 {\cal G}(z_{i},z_{j}) \equiv  \log E(z_{i},z_{j})
- \frac{1}{2}  \left(
\int_{z_{i}}^{z_{j}} \omega^{\mu} \right) ( 2 \pi {\mbox Im} \tau
 )_{\mu \nu}^{-1} \left( \int_{z_{i}}^{z_{j}} \omega^{\nu} 
\right)           \label{GFO}
\eeq

being the $g$-loop Green function \cite{GSW} that, also in this case, can be 
rescaled: 
\beq
 G(z_{i},z_{j}) \equiv  {\cal G}(z_{i},z_{j}) - \frac{1}{2}\log V'_{i}(0)
 V'_{j}(0)
\label{OGF}
\eeq

Let us consider first the simpler case of one-loop.
In this case the choice (\ref{GAUGE}) in (\ref{OGF}) reproduces the right 
translational invariant Green function and allows us to rewrite the amplitudes
in a very compact form:
\[
A_{M;1}\, =\,\frac{ \mbox{N Tr}[\lambda^{a_1}\cdots\lambda^{a_M}]}{(2\pi)^d(2\a')^{d/2}} (N_0^{ph.})^M
\int [ {\mbox d} m]^{1}_{M}
\]
\begin{equation}
\times \exp \bigg\{\frac{1}{2} \sum_{\stackrel{i,j=1}{i \neq j}}^{M} 
\left[ \sqrt{2 \a'} p^{(i)} + \xi_i  V'_i(0)
\partial_{z_{i}}\right]
\cdot \left[ \sqrt{2 \a'} p^{(j)} + \xi_j  V'_j(0)
\partial_{z_{j}} \right] G(z_{i},z_{j} ) \bigg\}
\label{VMP}
\end{equation}

Specializing the previous result for two external gluons we get \cite{CMPP1}:
\begin{equation}
A_{2;1} = N \frac{\mbox{Tr}[\lambda^{a_1}\lambda^{a_2}]}{(2\pi)^d(2\a')^{d/2}} 
({\cal N}^{ph.}_{0})^{2} 
\epsilon^{\mu} \epsilon^{\nu}
T_{\mu \nu}
\end{equation}
with
\[
T_{\mu \nu} = 2 \left( a_{1} \eta_{\mu \nu} + 4 \alpha^{'} a_{2} p^{(1)}_{\nu}
p^{(2)}_{\mu} \right)
\]
Again the explicit expression for the $a_i$`s can be found in Ref. 
\cite{CMPP1}.

By integrating by parts the terms containing the double derivative of 
$G(z_{i},z_{j})$,  we can rewrite the previous amplitude as follows
\cite{CMPP1}:
\begin{equation}
T_{\mu \nu} = 4 \alpha^{'} a_{2} \left[ p^{2} \eta_{\mu \nu} - p_{\nu} p_{\mu}
\right]
\end{equation}

This amplitude, as an off-shell string amplitude, is again transverse. 

In conclusion the choice (\ref{GAUGE}) for the local coordinates provides 
transverse one-loop two-massless amplitudes both in the open and 
closed bosonic string.
 
Let us now analyse how our procedure can be applied at tree level and for three
massless states (\ref{OMS}) of the open string \cite{LMPP}.

The starting point is again the relation (\ref{VMPO}) now specialized to $M=3$
and $g=0$. In this case it is simple to check that inserting in eq. (\ref{VMP})
the condition (\ref{GAUGE}), which we used at one-loop,  
one gets off-shell amplitudes that are not projective invariant.

Since the projective invariance of off-shell string amplitudes is an important 
requirement, being related to 
factorization and finiteness of the theory \cite{KV}, one has to require it
for the amplitude in exam. This implies that 
new expressions
for the functions $V'_i(0)$ and $V''_i(0)$ have to be considered.

In the specific case of the amplitude for three massless states
it turns out that in the limit $\alpha ^{\prime }\rightarrow 0$ 
this amplitude depends on the choice of $V_{i}$ only through the ratio
$V_{i}^{\prime \prime }(0)/(V_{i}^{\prime }(0))^{2}$. 
Requiring its being projective invariant in this limit allows
to select a family of functions of the punctures $(z_{1},z_{2},z_{3})$ 
depending on one parameter. The value of this latter can be fixed by 
requiring that the sum over all the anticyclic permutations of the lowest 
order term in
$\alpha'$, providing the field theory tree scattering amplitudes for
photons, be identically zero. In this way we univocally determine the
ratio $V_{i}^{\prime \prime }(0)/(V_{i}^{\prime}(0))^{2}$, which turns out to
be:

\begin{equation}
\frac{V_{i}^{\prime \prime }(0)}{(V_{i}^{\prime }(0))^{2}}=\frac{1}{\left(
z_{i}-z_{i+1}\right) }+\frac{1}{\left( z_{i}-z_{i-1}\right) }  \label{16}
\end{equation}

Furthermore, requiring that the whole amplitude, and not only
its low-energy limit, be projective invariant univocally fixes $V'_{i}(0)$
as:
\begin{equation}
V_{i}^{\prime }(0)=\frac{(z_{i}-z_{i+1})(z_{i}-z_{i-1})}{(z_{i+1}-z_{i-1})}%
\smallskip  \label{10}
\end{equation}
that corresponds to the so-called Lovelace choice,
where $V_{i}(z)$ is the projective transformation which maps the points
$\infty,0,1$ respectively in $z_{i-1},z_{i},z_{i+1}$.

By performing the above choices we get
the following projective invariant three-point off-shell open string
amplitude: 
\begin{eqnarray}
A_{0}^{3} &\simeq&g_d\varepsilon _{1\,\lambda }\varepsilon _{2\,\mu }\varepsilon
_{3\,\nu }\left\{ \left( \frac{\alpha ^{\prime }}{2}\right) \left[ \frac{1}{%
2}\left( p_{3}^{\lambda }-p_{2}^{\lambda }\right) \left( p_{1}^{\mu
}-p_{3}^{\mu }\right) \left( p_{2}^{\nu }-p_{1}^{\nu }\right) \right]
\right. +  \label{20} \\
&&+2\left. \left[ \eta ^{\lambda \mu }\left( p_{2}^{\nu }-p_{1}^{\nu
}\right) +\eta ^{\lambda \nu }\left( p_{1}^{\mu }-p_{3}^{\mu }\right) +\eta
^{\mu \nu }\left( p_{3}^{\lambda }-p_{2}^{\lambda }\right) \right] \right\} 
\nonumber
\end{eqnarray}

After having multiplied the amplitude (\ref{20}) by the Chan-Paton factor,
we get for the three-gluon amplitude \cite{LMPP}:

\begin{equation}
A_{0}^{3}(\mbox{gluons})\simeq g_d\,\,\varepsilon _{1\,\lambda }\varepsilon
_{2\,\mu }\varepsilon _{3\,\nu }f^{abc}\left\{ \eta ^{\lambda \mu }\left(
p_{1}^{\nu }-p_{2}^{\nu }\right) +\eta ^{\lambda \nu }\left( p_{3}^{\mu
}-p_{1}^{\mu }\right) +\eta ^{\mu \nu }\left( p_{2}^{\lambda
}-p_{3}^{\lambda }\right) \right\}.  \label{24}
\end{equation}

where $f$'s are the structure constants of $SU(\mbox{N})$.
This expression coincides with that of the three-gluon scattering
amplitude obtained in the usual covariant gauge.

\end{document}